\documentclass[journal]{IEEEtran}
\usepackage{float}

\usepackage{silence}
\ifCLASSINFOpdf
   \usepackage[pdftex]{graphicx} 
\else
   \usepackage[dvips]{graphicx}
   declare the path(s) where your graphic files are
  \graphicspath{{../eps/}}
  
\fi
\usepackage{subfig}
\usepackage{amsthm}
\usepackage[cmex10]{amsmath}
\usepackage{amssymb}
\usepackage{mathtools}

\usepackage{amsmath}       
\usepackage{mathtools}     
\usepackage{caption}
\usepackage{float}
\usepackage[table]{xcolor}
\usepackage{multirow}
\usepackage{multicol}
\usepackage{threeparttable}  
\usepackage{cuted}
\usepackage{enumerate}
\usepackage{enumitem}
\usepackage{epstopdf}
\usepackage{cite}
\usepackage{xcolor}

\def\BibTeX{{\rm B\kern-.05em{\sc i\kern-.025em b}\kern-.08em
    T\kern-.1667em\lower.7ex\hbox{E}\kern-.125emX}}

\usepackage{siunitx}
\usepackage{adjustbox}
\usepackage{array}
\newcolumntype{C}[1]{>{\centering\arraybackslash}m{#1}}
\usepackage{caption}
\usepackage{balance}
\usepackage{soul}
\definecolor{GreenForest}{rgb}{0.09, 0.45, 0.27}
\usepackage{algorithm}
\usepackage{algorithmic}

\usepackage{silence}
\usepackage[compatibility=false]{caption}

\usepackage{amsthm}

\begin{document}

\title{On the CSIT-Limited Delay Robustness of IR-HARQ in Finite Blocklength RSMA Networks}

\author{Shaima Abidrabbu
\thanks{S. Abidrabbu is with the IPR and License Agreements Department, Vestel Electronics, 45030 Manisa, Turkey. (e-mail: shaima.abidrabbu@vestel.com.tr).}
\protect}

\maketitle


\begin{abstract}
This letter studies the throughput and reliability of power-constrained rate-splitting multiple access (RSMA) with variable-length incremental-redundancy hybrid automatic repeat request (IR-HARQ) under finite blocklength (FBL) transmission and imperfect channel state information at the transmitter (CSIT). While perfect-CSIT feedback-delay tolerance approaches a natural ceiling as signal-to-noise ratio (SNR) grows, we show imperfect CSIT imposes a strictly lower, finite ceiling set solely by CSIT quality, one that transmit power cannot overcome. We derive analytical expressions and tight approximations for common- and private-stream decoding failure probabilities, capturing FBL and CSIT uncertainty via a unified CSIT-degradation factor. Also, we show that residual interference saturates effective stream power at high SNR, limiting reliability and producing this delay-tolerance ceiling. Simulations validate the analysis and quantify the ceiling, offering design insights for delay-robust RSMA under imperfect CSIT.
\end{abstract}

\begin{IEEEkeywords}
RSMA, HARQ, finite block-length coding.  
\end{IEEEkeywords}

\section{Introduction}
Hybrid automatic repeat request (HARQ) improves transmission reliability by accumulating redundancy across retransmissions, making it well-suited to finite-blocklength (FBL) transmission with stringent reliability and latency requirements. Rate-splitting multiple access (RSMA), in turn, provides flexible interference management and has been extensively studied in FBL systems for ultra-reliable low-latency communication (uRLLC) applications~\cite{1,2,4,5}. When these techniques are combined, however, the reliability benefits of retransmission depend critically on the quality and timeliness of channel state information at the transmitter (CSIT). Imperfect CSIT leaves residual multi-user interference, while feedback delay postpones the use of updated retransmission information. Their interaction can therefore fundamentally limit the reliability and delay robustness of HARQ-assisted~RSMA.

Recent FBL-RSMA studies have incorporated practical impairments such as imperfect CSIT, successive interference cancellation (SIC) errors, and hardware impairments, demonstrating their adverse impact on reliability~\cite{swipt_fbl,singh_uav,autonomous_fbl}. These studies, however, primarily consider single-shot transmission. HARQ has also been investigated in RSMA through adaptive switching among RSMA, non-orthogonal multiple access (NOMA), and space division multiple access (SDMA)~\cite{13,15}, common-stream embedding of private-data redundancy~\cite{14}, and uplink HARQ-RSMA with user-dependent retransmission intervals~\cite{10557543}. More recently, variable-length incremental-redundancy HARQ (IR-HARQ) for FBL-RSMA has been studied with adaptive sub-codeword allocation and feedback-delay-aware throughput conditions~\cite{11357511}. The study establishes the throughput and delay benefits of variable-length IR-HARQ under perfect CSIT, while the impact of imperfect CSIT remains uncharacterized.

CSIT uncertainty can fundamentally limit these benefits. Under imperfect CSIT, residual interference scales with transmit power, causing the effective stream power to saturate at high SNR. Consequently, further power increases provide diminishing reliability gains and cannot fully overcome feedback-delay limitations, resulting in a finite delay-tolerance ceiling determined by CSIT quality. Motivated by this limitation, we develop a CSIT-aware variable-length IR-HARQ framework for FBL-RSMA that jointly captures CSIT estimation error, adaptive sub-codeword allocation, and feedback delay. We derive the decoding-failure probabilities and characterize their high-SNR behavior, revealing a finite upper bound on feedback-delay tolerance that cannot be overcome by increasing transmit power or the number of redundancy rounds. The main contributions are summarized as follows:
\begin{itemize}[leftmargin=*]
\item We characterize the impact of imperfect CSIT on variable-length IR-HARQ RSMA and prove that the maximum tolerable relative feedback delay converges to a finite CSIT-dependent ceiling, strictly below the perfect-CSIT bound $(M-1)/M$ for nonzero CSIT error.

\item We identify the underlying mechanism: CSIT-induced residual interference causes the effective stream power to saturate at high SNR, which simultaneously limits decoding reliability and delay robustness.

\item We derive analytical expressions and tight high-SNR approximations for common- and private-stream decoding failures, incorporating FBL effects and CSIT uncertainty through a unified CSIT-degradation factor.

\item We derive a sufficient condition for IR-HARQ to outperform no-HARQ transmission in the presence of both imperfect CSIT and feedback delay, thereby establishing the conditions under which retransmission remains beneficial in the CSIT-limited regime.
\end{itemize}

\section{System Model Under Imperfect CSIT}
We consider the redundancy-version-embedded RSMA IR-HARQ framework of~\cite{14}, where a base station (BS) equipped with $N_t$ transmit antennas serves $K$ single-antenna users with $N_t\ge K+1$. The redundancy versions of all users are multiplexed into the common stream, while the remaining information is transmitted through the private streams. The channel at the BS is given as follows 
\begin{equation}
\mathbf{h}_k=\hat{\mathbf{h}}_k+\mathbf{e}_k,
\label{eq:csit_error}
\end{equation}
where $\hat{\mathbf{h}}_k$ is the estimated channel and $\mathbf{e}_k\sim\mathcal{CN}(\mathbf{0},\sigma_e^2\mathbf{I}_{N_t}$ is the estimation error, independent across users and HARQ rounds. The parameter $\sigma_e^2\in[0,1)$ characterizes the CSIT quality, with $\sigma_e^2=0$ corresponding to perfect CSIT. The BS designs conventional zero-forcing (ZF) precoders based on the estimated channels without robust optimization. We further assume identical CSIT quality for all users, perfect CSIR, and error-free HARQ feedback. Therefore, the transmitted signal can be written as
\begin{equation}
\mathbf{x}=\hat{\mathbf{p}}_cs_c+\sum_{k=1}^{K}\hat{\mathbf{p}}_ks_k,
\end{equation}
where $\hat{\mathbf{P}}=[\hat{\mathbf{p}}_c,\hat{\mathbf{p}}_1,\ldots,\hat{\mathbf{p}}_K]$ denotes the ZF precoder computed from $\{\hat{\mathbf{h}}_k\}$. At the $m$th HARQ round, user $k$ receives the following signal 
\begin{equation}
y_k^{(m)}=\mathbf{h}_k^H\mathbf{x}^{(m)}+n_k^{(m)},
\label{eq:received}
\end{equation}
where $n_k^{(m)}$ is additive white Gaussian noise and it is given as $n_k^{(m)}\sim\mathcal{CN}(0,\sigma_k^2)$. Using \eqref{eq:csit_error}, \eqref{eq:received} can be given as 
\begin{equation}
y_k^{(m)}=\hat{\mathbf{h}}_k^H\mathbf{x}^{(m)}+\mathbf{e}_k^H\mathbf{x}^{(m)}+n_k^{(m)},
\label{eq:received_split}
\end{equation}
where the second term represents the residual interference caused by CSIT uncertainty. Since the ZF precoder is designed using $\{\hat{\mathbf{h}}_k\}$, multiuser interference is cancelled only with respect to the estimated channels.

Conditioned on the estimated channels, the residual interference is zero-mean with variance that can be written as follows
\begin{equation}
\mathbb{E}\!\left[
\left|
\mathbf{e}_k^H\mathbf{x}
\right|^2
\right]
=
\sigma_e^2
\left(
\|\hat{\mathbf{p}}_c\|^2P_c
+
\sum_{j=1}^{K}
\|\hat{\mathbf{p}}_j\|^2P_j
\right).
\end{equation}
Following the conventional imperfect-CSIT model, this interference is treated as Gaussian noise, yielding the effective noise variance to be expressed as follows 
\begin{equation}
\tilde{\sigma}_k^2
=
\sigma_k^2
+
\sigma_e^2
\left(
\|\hat{\mathbf{p}}_c\|^2P_c
+
\sum_{j=1}^{K}
\|\hat{\mathbf{p}}_j\|^2P_j
\right),
\label{eq:effective_noise}
\end{equation}
which replaces $\sigma_k^2$ in the FBL decoding analysis. Each user first decodes the common stream by treating the private streams as interference. After successful SIC, the private stream is decoded under the effective noise variance in~\eqref{eq:effective_noise}. If decoding fails, additional redundant versions are transmitted and combined with the previously received packets according to the IR-HARQ protocol.

\section{ANALYSIS OF THE PROPOSED APPROACH}
This section develops the throughput and reliability analysis for the proposed scheme under imperfect CSIT, establishing FBL rate and outage expressions, an optimized power-allocation structure, and a joint sufficient condition characterizing when IR-HARQ improves throughput under the combined effect of CSIT imperfection and feedback delay.

\subsection{FBL Rate and Reliability Model Under Imperfect CSIT}
Given the channel decomposition $\mathbf{h}_k$ with $\mathbf{e}_k \sim \mathcal{CN}(\mathbf{0},\sigma_e^2\mathbf{I}_{N_t})$, the residual interference term identified in the reception model contributes an additive power $\sigma_e^2 P_{\mathrm{total}}$ to the noise floor seen by each user. For a fixed interference power, Gaussian interference is the worst case over all interference distributions with matching second moment, treating this residual term as Gaussian yields a rigorous lower bound on the achievable rate, which we adopt as the design basis for reliability-oriented resource allocation. This gives an effective noise variance as follows
\begin{equation}
    \tilde{\sigma}_k^2 = \sigma_k^2 + \sigma_e^2 P_{\mathrm{total}},
    \label{eq:noise_eff}
\end{equation}
and an equivalent CSIT-degradation factor as $\kappa \triangleq \tilde{\sigma}_k^2/\sigma_k^2 = 1 + \sigma_e^2 P_{\mathrm{total}}/\sigma_k^2 \geq 1$, so that the effective transmit power available to stream $x \in \{c,k\}$ is $\tilde{P}_x = P_x/\kappa$.

The achievable FBL rate and target error probability of stream $x$ then follow the non-asymptotic coding-rate characterization under quasi-static fading~\cite{6802432}, evaluated at the effective power $\tilde P_x$ as follows: 
{\small\begin{align}
    R_x(L_x, \tilde P_x, \delta_x) &\approx \log(1+g\tilde P_x) - \sqrt{\tfrac{V_x(\tilde P_x)}{L_x}}\,Q^{-1}(\delta_x) + F_x, \label{eq:rate_csit}\\
    \delta_x(L_x, R_x, \tilde P_x) &\approx \mathbb{E}\!\left[Q\!\left(\tfrac{\sqrt{L_x}\left[\log(1+g\tilde P_x)-R_x\right]}{\sqrt{1-(1+g\tilde P_x)^{-2}}}\right)\right], \label{eq:err_csit}
\end{align}
}where $V_x(\tilde P_x) = 1-(1+g\tilde P_x)^{-2}$ is the channel dispersion and $F_x = O(\log L_x/L_x)$ is the third-order term. Expanding \eqref{eq:rate_csit} to first order in $\sigma_e^2$ gives the CSIT-induced rate penalty
\begin{equation}
    \Delta R_x(\sigma_e^2) \approx \frac{gP_x}{1+gP_x}\log_2(e)\,\frac{\sigma_e^2 P_{\mathrm{total}}}{\sigma_k^2} + O(\sigma_e^4),
    \label{eq:rate_loss}
\end{equation}
which shows that the rate penalty scales with the high-SNR efficiency factor $gP_x/(1+gP_x)$: streams operating at higher effective SNR are disproportionately more sensitive to CSIT imperfection. This observation directly informs the power-split decision in Section III-D.

\subsection{Throughput Expression}
The throughput of the proposed scheme can be written as
\begin{equation}
    K_{\mathrm{RSMA}} = K_c\big(1-\Omega_c(\kappa)\big) + \sum_{k=1}^K K_{p,k}\big(1-\Omega_{p,k}(\kappa)\big),
    \label{eq:throughput}
\end{equation}
where $K_c$ and $K_{p,k}$ are the information payloads of the common and $k$-th private streams, and $\Omega_c(\kappa), \Omega_{p,k}(\kappa)$ are their cumulative decoding failure probabilities under CSIT-degradation factor $\kappa$, derived in Section III-C. The expected total channel uses over $M$ HARQ rounds, incorporating feedback delay $d$, is
\begin{equation}
    T = \sum_{m=1}^M \Omega_{m-1}(\kappa)\,l_{\mathrm{total}}(m) + d\sum_{m=1}^{M-1}\Omega_{m-1}(\kappa),
    \label{eq:T}
\end{equation}
with $l_{\mathrm{total}}(m) = l_c(m)+\sum_k l_{p,k}(m)$ the sub-codeword length used in round $m$. The impact of CSIT imperfections on $\eta_{\mathrm{RSMA}}$ is captured entirely by $\Omega_c(\kappa)$ and $\Omega_{p,k}(\kappa)$, while the underlying throughput structure remains unchanged, with the same successive common-then-private decoding order and blocklength.

\subsection{Decoding Failure Probabilities Under Imperfect CSIT}
The decoding failure probabilities and their tight bounds are derived, where the central reliability metric through which CSIT uncertainty propagates into throughput.
For a stream $x\in\{c,k\}$ at the end of round $m$, the decoding failure probability under FBL coding and effective power $\tilde P_x$ is then given as 
{\small\begin{equation}
    \Omega_x(\kappa) = \mathbb{E}_g\!\left[Q\!\left(\frac{\sqrt{l_{x,(M-1)}}\left[\log(1+g\tilde{P}_x)-R_{x,(M-1)}\right]}{\sqrt{1-(1+g\tilde{P}_x)^{-2}}}\right)\right].
    \label{eq:outage_def}
\end{equation}
}Under quasi-static Rayleigh fading with $g=z\sim\mathrm{Exp}(1)$,
\begin{equation}
\begin{aligned}
\Omega_x(\kappa)
= \int_0^\infty e^{-z}\,
Q\!\Biggl(
&\frac{\sqrt{l_{x,(M-1)}}}
{\sqrt{1-(1+\tilde{P}_x z)^{-2}}} \\
&\times
\left[\log(1+\tilde{P}_x z)-R_{x,(M-1)}\right]
\Biggr)\,dz,
\end{aligned}
\label{eq:outage_int}
\end{equation}
which we evaluate numerically via Gauss--Laguerre quadrature. For analytical insight, tight bounds follow by first-order expansion of the $Q$-function argument about $z^\star(\kappa) = \kappa\left(e^{R_{x,(M-1)}}-1\right)/P_x$ (lower bound), and the exponential tail bound $Q(u)\le \tfrac12 e^{-u^2/2}$ combined with properties of the incomplete Gamma function (upper bound).

\emph{Lemma 1 (Bound-Gap Widening Under CSIT Imperfection):} Let $\Omega_x^{\mathrm{LB}}(\kappa), \Omega_x^{\mathrm{UB}}(\kappa)$ denote the bounds obtained above. Then both bounds are non-decreasing in $\kappa$, and the bound gap $\Delta(\kappa) \triangleq \Omega_x^{\mathrm{UB}}(\kappa)-\Omega_x^{\mathrm{LB}}(\kappa)$ is non-decreasing for $\kappa \geq 1$. Furthermore, the outage sensitivity to CSIT quality is
\begin{equation}
    \left.\frac{\partial \Omega_x}{\partial \sigma_e^2}\right|_{\sigma_e^2=0} = -\frac{P_xP_{\mathrm{total}}}{\sigma_k^2}\left.\frac{\partial \Omega_x}{\partial P_x}\right|_{\kappa=1}.
    \label{eq:sensitivity}
\end{equation}

\emph{Proof:} Since $\tilde P_x = P_x/\kappa$ is strictly decreasing in $\kappa$, and $Q(\cdot)$ is monotone in its argument, $z^\star(\kappa)$ shifts monotonically and both bound constructions inherit monotonicity in $\kappa$ term-by-term. Differentiating $\tilde P_x$ with respect to $\sigma_e^2$ at $\sigma_e^2=0$ gives $\partial \tilde P_x/\partial \sigma_e^2 = -P_xP_{\mathrm{total}}/\sigma_k^2$; applying the chain rule to \eqref{eq:outage_int} yields \eqref{eq:sensitivity}. $\blacksquare$

Lemma 1 shows that CSIT imperfection systematically widens the outage-bound gap and quantifies the approximation slack introduced by imperfect CSIT. Meanwhile, \eqref{eq:sensitivity} provides a closed-form gradient that reuses the existing $\partial\Omega_x/\partial P_x$ term required for power-allocation optimization, without additional computational cost.

\subsection{CSIT-Aware Power Allocation and Asymptotic Saturation Analysis} 
To maximize network performance under joint impairments, the optimal power allocation is formulated via the Lagrangian 
\begin{equation}
    \mathcal{L} = \frac{K_{\mathrm{RSMA}}}{T} - \lambda\Big(P_c+\sum_{k=1}^K P_k - P_{\mathrm{total}}\Big),
    \label{eq:lagrangian}
\end{equation}
where stationarity conditions $\partial\mathcal{L}/\partial P_c=0$ and $\partial\mathcal{L}/\partial P_k=0$ are evaluated using $\partial\Omega_x/\partial P_x$ from \eqref{eq:sensitivity}, alternated iteratively with sub-codeword length updates. However, because higher effective SNR streams exhibit heightened sensitivity to estimation errors \eqref{eq:rate_loss}, optimization convergence is accelerated by initializing the common-stream power allocation fraction via a CSIT-aware heuristic $\alpha_c = \alpha_0(1 + \gamma \sigma_e^2 P_{\mathrm{total}}/\sigma_k^2)$, where $\alpha_0$ is the baseline split and $\gamma > 0$ controls error sensitivity.

This adaptive splitting is structurally crucial because increasing $P_{\mathrm{total}}$ does not monotonically improve reliability under imperfect CSIT. Specifically, evaluating the effective power $\tilde P_x = P_x/\kappa$ reveals that 
{\small\begin{equation}
    \frac{\partial \tilde P_x}{\partial P_{\mathrm{total}}} = -\frac{P_x\sigma_e^2/\sigma_k^2}{\kappa^2} \;\xrightarrow{P_{\mathrm{total}}\to\infty}\; 0, \quad \tilde P_x \to \frac{P_x}{P_{\mathrm{total}}}\cdot\frac{\sigma_k^2}{\sigma_e^2},
    \label{eq:saturation}
\end{equation}
}i.e., $\tilde P_x$ saturates to a finite ceiling rather than growing unboundedly with $P_{\mathrm{total}}$. This CSIT-limited saturation, analogous to SINR ceilings in interference-limited MIMO systems, implies that beyond a critical $P_{\mathrm{total}}$ threshold, further power investment yields vanishing reliability gains. Consequently, the gradients $\partial\Omega_x/\partial P_x$ vanish, causing the marginal benefit of the Lagrangian search to collapse and forcing the system into a rigid, CSIT-bounded throughput regime. This power saturation directly dictates the high-SNR behavior of the delay-robustness condition established next.

\subsection{Joint CSIT-Delay Robustness of IR-HARQ RSMA}
We characterize the throughput gain of IR-HARQ over no-HARQ transmission under imperfect CSIT and feedback delay $d$, with normalized delay $d_{\mathrm{rel}}=d/T$.

\emph{Theorem 1 (Joint Sufficient Condition for HARQ Gain):} Consider an RSMA system with $M$ HARQ rounds, sub-codeword lengths $\{l_c(m),l_{p,k}(m)\}$, and cumulative decoding failure probabilities $\{\Omega_{m-1}(\kappa)\}$ under CSIT-degradation factor $\kappa \geq 1$. For variable-length coding, IR-HARQ yields a throughput gain over no-HARQ if
\begin{equation}
\begin{aligned}
d_{\mathrm{rel}}
\leq r(\kappa)
&\triangleq
\min_{\{l_c(m),\,l_{p,k}(m)\}} \\
&\frac{
l_{\mathrm{total}}(M)
-
\sum_{m=1}^M
\Omega_{m-1}(\kappa)\,
l_{\mathrm{total}}(m)
}{
T\displaystyle\sum_{m=1}^{M-1}
\Omega_{m-1}(\kappa)
}.
\end{aligned}
\label{eq:theorem1}
\end{equation}
Violation of \eqref{eq:theorem1} implies $\eta_{\mathrm{HARQ}} \leq \eta_{\mathrm{no\text{-}HARQ}}$ under CSIT quality $\sigma_e^2$.

\emph{Proof:} For HARQ, the expected total channel uses are $T_{\mathrm{HARQ}} = \sum_{m=1}^M \Omega_{m-1}(\kappa)\,l_{\mathrm{total}}(m) + d_{\mathrm{rel}}T\sum_{m=1}^{M-1}\Omega_{m-1}(\kappa)$; for no-HARQ ($M=1$), $T_{\mathrm{no\text{-}HARQ}} = l_{\mathrm{total}}(1)$. Imposing $T_{\mathrm{HARQ}} \leq T_{\mathrm{no\text{-}HARQ}}$ and rearranging for $d_{\mathrm{rel}}$ gives \eqref{eq:theorem1} directly; the minimization over $\{l_c(m),l_{p,k}(m)\}$ follows by optimizing the sub-codeword allocation subject to $\sum_m l_x(m) = T_x$, which yields the closed-form allocation
\begin{equation}
    l_x^\star(m) = T_x\,\frac{\Omega_{m-1}(\kappa)}{\sum_{j=1}^M \Omega_{j-1}(\kappa)},
    \label{eq:opt_alloc}
\end{equation}
obtained via Lagrangian minimization of $\sum_m \Omega_{m-1}(\kappa)l_{\mathrm{total}}(m)$ subject to the length-budget constraint, with the Cauchy--Schwarz inequality confirming this allocation is optimal. $\blacksquare$

\emph{Corollary 1 (Delay-Robustness Ceiling Under Imperfect CSIT):} Since $\Omega_0 = 1$ irrespective of channel quality (the first HARQ round is always transmitted), the delay-robustness margin $r(\kappa)$ is bounded above by a \emph{natural ceiling} $(M-1)/M$ even under perfect CSIT: as $\mathrm{SNR}\to\infty$ with $\sigma_e^2=0$, $\Omega_1,\ldots,\Omega_{M-1}\to 0$ while $\Omega_0=1$ remains fixed, giving
\begin{equation}
    \lim_{\mathrm{SNR}\to\infty} r(1) = \frac{1-\frac{1}{M}\cdot 1}{1} = \frac{M-1}{M}.
    \label{eq:natural_ceiling}
\end{equation}
Under imperfect CSIT ($\sigma_e^2 > 0$), the saturation in \eqref{eq:saturation} implies $\tilde P_x \to (P_x/P_{\mathrm{total}})(\sigma_k^2/\sigma_e^2)$, a finite effective power, so $\Omega_1,\ldots,\Omega_{M-1}$ approach strictly positive error floors $\Omega_{m}^{\infty}(\sigma_e^2) > 0$ rather than vanishing. Consequently, $r(\kappa)$ saturates to a ceiling \emph{strictly below} the natural bound:
\begin{equation}
\begin{aligned}
\lim_{\mathrm{SNR}\to\infty} r(\kappa)
&=
\frac{
1-\frac{1}{M}\left(1+\sum_{m=1}^{M-1}
\Omega_{m}^{\infty}(\sigma_e^2)\right)
}{
1+\sum_{m=1}^{M-2}\Omega_{m}^{\infty}(\sigma_e^2)
} \\
&\triangleq r_\infty(\sigma_e^2)
< \frac{M-1}{M}.
\end{aligned}
\label{eq:r_infinity}
\end{equation}

\begin{figure*}[ht]
    \centering
    \includegraphics[width=\linewidth]{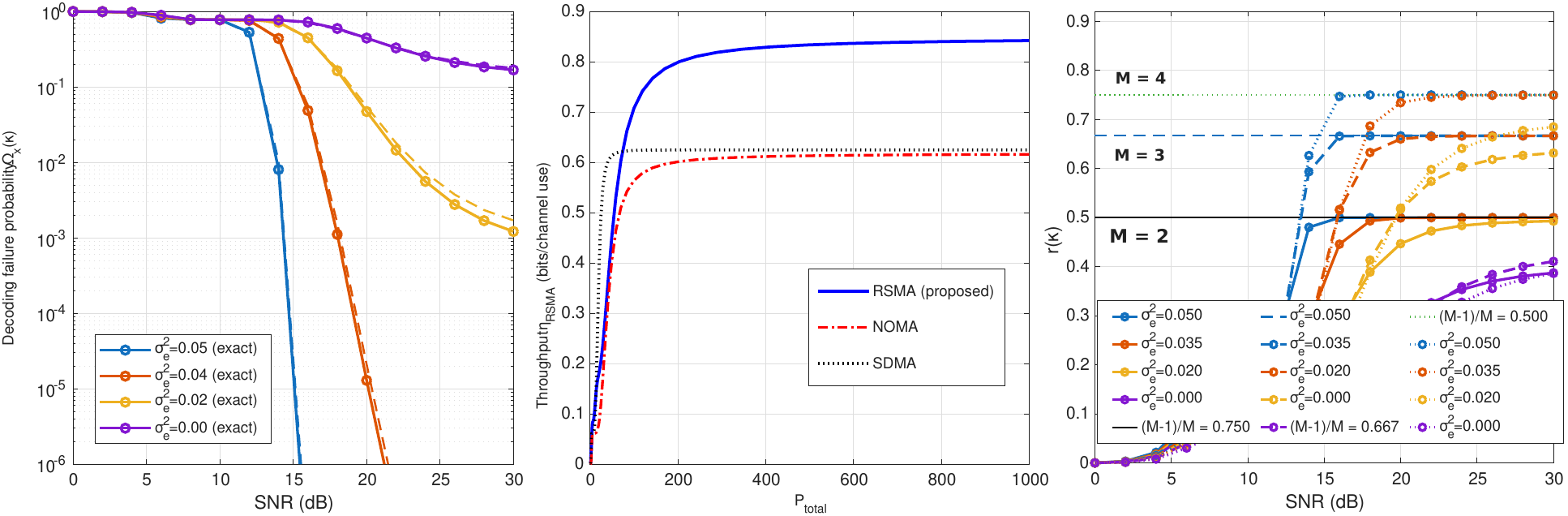}
     \caption{Impact of imperfect CSIT on RSMA IR-HARQ for $\sigma_e^2\in\{0,0.02,0.035,0.05\}$: (a) decoding failure probability $\Omega_x(\kappa)$ versus SNR, illustrating CSIT-induced error floors; (b) throughput versus total transmit power for $\sigma_e^2=0.035$, comparing RSMA, NOMA, and SDMA; (c) maximum tolerable relative feedback delay $r(\kappa)$ versus SNR for different numbers of users, illustrating the impact of CSIT errors relative to the natural bound $(M-1)/M$.}
    \label{fig:placeholder}
\end{figure*} 

\emph{Proof:} Substituting the saturated $\tilde P_x$ into \eqref{eq:outage_int} shows the integrand no longer collapses to zero as $z\to\infty$-scaling with $\tilde P_x\to\infty$; instead, $\tilde P_x$ is bounded above by $(P_x/P_{\mathrm{total}})(\sigma_k^2/\sigma_e^2)$ uniformly in SNR, so each $\Omega_m$, $m\geq1$, is bounded below by its value at this ceiling power, establishing $\Omega_m^{\infty}(\sigma_e^2)>0$ for any $\sigma_e^2>0$. Substituting these floors into the definition of $r(\kappa)$ and taking the SNR limit yields \eqref{eq:r_infinity}; since each $\Omega_m^\infty(\sigma_e^2) > 0$ strictly increases the denominator and strictly decreases the numerator relative to the $\sigma_e^2=0$ case in \eqref{eq:natural_ceiling}, $r_\infty(\sigma_e^2) < (M-1)/M$ strictly. $\blacksquare$

Corollary 1 refines the delay robustness picture from two complementary perspectives. First, the number of HARQ rounds $M$ imposes a fundamental, CSIT independent ceiling of $(M-1)/M$ on the achievable delay tolerance. This bound cannot be overcome by simply increasing the transmit power or improving the channel quality, since round 1 is unconditionally transmitted regardless of SNR. Second, and more importantly for this letter, imperfect CSIT lowers the achievable ceiling to $r_\infty(\sigma_e^2)$, which is strictly below $(M-1)/M$ for nonzero $\sigma_e^2$. The resulting gap is determined by the CSIT error variance $\sigma_e^2$ and cannot be eliminated through additional transmit power. Keeping $\sigma_e^2$ constant as $\mathrm{SNR}\rightarrow\infty$ represents a practical regime in which CSIT degradation is dominated by feedback delay or Doppler-induced channel aging due to user mobility, rather than by pilot estimation noise. This observation leads to the central design implication of this letter: transmit power and CSIT quality are not interchangeable means of improving delay robustness in IR-HARQ RSMA. When CSIT accuracy is bounded, system designers must therefore budget the feedback delay according to the CSIT-limited ceiling $r_\infty(\sigma_e^2)$ rather than the more optimistic CSIT-independent ceiling $(M-1)/M$, which is attainable only under perfect CSIT.

\section{Simulation Results}
This section evaluates the proposed scheme over quasi-static Rayleigh fading channels with FBL constraints and imperfect CSIT. We consider $\sigma_e^2=0.02$ and $0.05$ to represent low- and high-mobility 5G scenarios, respectively~\cite{ref_5G_mobility}. Decoding failure probability, throughput, and delay-robustness margin are obtained from the closed-form expressions in Section III, while \eqref{eq:outage_int} is evaluated using Gauss--Laguerre quadrature. Simulation parameters are summarized in Table~I.

\begin{table}[ht]
\centering
\caption{Simulation Parameters}
\label{tab:sim_params}
\setlength{\tabcolsep}{9pt}
\renewcommand{\arraystretch}{1.5}
\begin{tabular}{|l|c|}
\hline
\textcolor{black}{\textbf{Parameter}} & \textcolor{black}{\textbf{Value}} \\
\hline
\textcolor{black}{Number of users} & \textcolor{black}{$K=3$} \\
\hline
\textcolor{black}{BS antennas} & \textcolor{black}{$N_t=4$} \\
\hline
\textcolor{black}{Noise variance} & \textcolor{black}{$\sigma_k^2=1$} \\
\hline
\textcolor{black}{Common-stream power fraction} & \textcolor{black}{$\alpha_c=0.4$} \\
\hline
\textcolor{black}{Target rate} & \textcolor{black}{$R_x=1.0$ bits/channel use} \\
\hline
\textcolor{black}{Per-round blocklength} & \textcolor{black}{$l_0=25$} \\
\hline
\textcolor{black}{HARQ rounds} & \textcolor{black}{$M=3$} \\
\hline
\textcolor{black}{CSIT error variance} & \textcolor{black}{$\sigma_e^2\in\{0,0.02,0.035,0.05\}$} \\
\hline
\textcolor{black}{SNR range} & \textcolor{black}{$0$--$30$ dB (2 dB steps)} \\
\hline
\textcolor{black}{Relative feedback delay} & \textcolor{black}{$d_{\mathrm{rel}}=0.05$} \\
\hline
\textcolor{black}{Blocklength budget} & \textcolor{black}{$T\in\{600\}$} \\
\hline
\end{tabular}
\end{table}

Fig.~1(a) shows the decoding failure probability $\Omega_x(\kappa)$ versus SNR for different CSIT-error levels. The Gauss--Laguerre evaluation of \eqref{eq:outage_int} closely follows the high-SNR approximation, supporting the tightness of Lemma~1. With perfect CSIT, $\Omega_x(\kappa)$ continues to decrease with SNR, whereas any $\sigma_e^2>0$ leads to an error floor at high SNR. This floor results from the saturation of the effective power $\tilde P_x$ in \eqref{eq:saturation}, demonstrating that increasing transmit power cannot eliminate the reliability loss caused by imperfect CSIT.

Fig.~1(b) compares the resulting throughput with NOMA and SDMA for $\sigma_e^2=0.035$. All schemes exhibit high-SNR saturation, reflecting the common CSIT-induced limitation. RSMA, however, achieves the highest throughput plateau, approximately $35\%$ above NOMA and SDMA, owing to the additional payload enabled by its common stream. Thus, RSMA retains a throughput advantage even when all schemes experience the same CSIT degradation.

Fig.~1(c) illustrates the maximum tolerable relative feedback delay $r(\kappa)$ for $M\in\{2,3,4\}$. In contrast to the perfect-CSIT limit $(M-1)/M$, imperfect CSIT causes $r(\kappa)$ to saturate at a finite value determined by the CSIT-error level and the number of users. The ceiling decreases as the CSIT error becomes more pronounced, while its dependence on $M$ follows the natural $(M-1)/M$ scaling. These results confirm that CSIT quality directly limits the delay robustness of RSMA-IR-HARQ, particularly in high-mobility regimes.

\section{Conclusion and Future Work}
This letter investigated variable-length IR-HARQ RSMA under FBL and imperfect CSIT. We showed that CSIT errors induce a power-saturation regime that limits decoding reliability and imposes a finite upper bound on the maximum tolerable feedback delay, strictly below $(M-1)/M$. This saturation establishes a common fundamental limit on reliability, throughput, and delay tolerance: increasing transmit power cannot overcome these bounds, whereas improved CSIT quality can. These results highlight the need to design practical RSMA systems with delay budgets commensurate with achievable CSIT accuracy. Future work will consider robust precoding to mitigate the CSIT-induced ceiling, imperfect SIC, and time-varying channels with adaptive feedback scheduling.

\bibliographystyle{IEEEtran}
\bibliography{IEEEabrv.bib,ref.bib}{}

\begin{thebibliography}{10}
\providecommand{\url}[1]{#1}
\csname url@samestyle\endcsname
\providecommand{\newblock}{\relax}
\providecommand{\bibinfo}[2]{#2}
\providecommand{\BIBentrySTDinterwordspacing}{\spaceskip=0pt\relax}
\providecommand{\BIBentryALTinterwordstretchfactor}{4}
\providecommand{\BIBentryALTinterwordspacing}{\spaceskip=\fontdimen2\font plus
\BIBentryALTinterwordstretchfactor\fontdimen3\font minus \fontdimen4\font\relax}
\providecommand{\BIBforeignlanguage}[2]{{%
\expandafter\ifx\csname l@#1\endcsname\relax
\typeout{** WARNING: IEEEtran.bst: No hyphenation pattern has been}%
\typeout{** loaded for the language `#1'. Using the pattern for}%
\typeout{** the default language instead.}%
\else
\language=\csname l@#1\endcsname
\fi
#2}}
\providecommand{\BIBdecl}{\relax}
\BIBdecl

\bibitem{1}
J.~Zhu \emph{et~al.}, ``Joint optimization of user scheduling, rate allocation, and beamforming for {RSMA} finite blocklength transmission,'' \emph{IEEE Internet of Things J.}, vol.~11, no.~17, pp. 27\,904--27\,915, Sep. 2024.

\bibitem{2}
Y.~Xu, Y.~Mao, O.~Dizdar, and B.~Clerckx, ``{Max-Min} fairness of rate-splitting multiple access with finite blocklength communications,'' \emph{IEEE Trans. Veh. Technol.}, vol.~72, no.~5, pp. 6816--6821, May 2023.

\bibitem{4}
S.~Pala \emph{et~al.}, ``Spectral-efficient {RIS-A}ided {RSMA URLLC}: Toward mobile broadband reliable low latency communication ({mBRLLC}) system,'' \emph{IEEE Trans. Wireless Commun.}, vol.~23, no.~4, pp. 3507--3524, Apr. 2024.

\bibitem{5}
S.~K. Singh \emph{et~al.}, ``{RSMA} for hybrid {RIS-UAV}-aided full-duplex communications with finite blocklength codes under imperfect {SIC},'' \emph{IEEE Trans. Wireless Commun.}, vol.~22, no.~9, pp. 5957--5975, Sep. 2023.

\bibitem{swipt_fbl}
F.~Karim \emph{et~al.}, ``Finite blocklength analysis for {SWIPT}-enabled {RSMA} networks under realistic assumptions,'' \emph{IEEE Trans. Wireless Commun.}, vol.~24, no.~9, pp. 8014--8024, 2025.

\bibitem{singh_uav}
S.~K. Singh \emph{et~al.}, ``Performance analysis and optimization of {RSMA} enabled {UAV}-aided {IBL} and {FBL} communication with imperfect {SIC} and {CSI},'' \emph{IEEE Trans. Wireless Commun.}, vol.~22, no.~6, pp. 3714--3732, 2023.

\bibitem{autonomous_fbl}
Y.~Wang \emph{et~al.}, ``{Autonomous Driving With RSMA-Enabled Finite Blocklength Transmissions: Ergodic Performance Analysis and Optimization},'' \emph{IEEE Trans. Wireless Commun.}, vol.~25, pp. 2954--2969, 2026.

\bibitem{13}
S.~Abidrabbu, S.~R. Ali, and H.~Arslan, ``A novel {HARQ} design for {RSMA} networks,'' \emph{IEEE Internet of Things J.}, pp. 1--1, 2023.

\bibitem{15}
S.~Abidrabbu and H.~Arslan, ``A new flexible {HARQ} assisted by rate splitting multiple access,'' \emph{IEEE Trans. Netw. Sci. Eng. (Early Access)}, pp. 1--13, 2025.

\bibitem{14}
S.~Abidrabbu, S.~R. Ali, H.~M. Furqan, and H.~Arslan, ``An {IR-HARQ} based {RSMA} for reliable and low-latency communication,'' \emph{IEEE Trans. Green Commun. Netw.}, vol.~8, no.~4, pp. 1308--1317, Dec. 2024.

\bibitem{10557543}
Y.~Liu, B.~Clerckx, and P.~Popovski, ``Performance analysis of uplink rate-splitting multiple access with hybrid {ARQ},'' \emph{IEEE Trans. Wireless Commun.}, vol.~23, no.~10, pp. 14\,201--14\,214, Oct. 2024.

\bibitem{11357511}
S.~Abidrabbu and H.~Arslan, ``{Variable-Length IR-HARQ for Delay-Robust Throughput in Finite Blocklength RSMA Networks},'' \emph{IEEE Wireless Commun. Lett.}, vol.~15, pp. 1449--1453, 2026.

\bibitem{6802432}
W.~Yang, G.~Durisi, T.~Koch, and Y.~Polyanskiy, ``Quasi-static multiple-antenna fading channels at finite blocklength,'' \emph{IEEE Trans. Inf. Theory}, vol.~60, no.~7, pp. 4232--4265, July 2014.

\bibitem{ref_5G_mobility}
{3GPP}, ``{Study on channel model for frequencies from 0.5 to 100 GHz},'' {3rd Generation Partnership Project (3GPP)}, Tech. Rep. {TR 38.901 V17.0.0}, 2022.

\end{thebibliography}

\end{document}